\documentclass[prb,twocolumn,showpacs]{revtex4}
\usepackage{color,soul}
\usepackage{amsmath}
\usepackage{dcolumn}
\usepackage{graphicx}
\usepackage{bm}
\usepackage{amssymb}
\usepackage{subfigure}
\begin{document}

\title{Why the Schwinger Boson mean field theory fails to describe the spin dynamics of the triangular lattice antiferromagnetic Heisenberg model?}
\author{Qiu Zhang and Tao Li}
\affiliation{Department of Physics, Renmin University of China, Beijing 100872, P.R.China}

\begin{abstract}
We find that the Schwinger Boson mean field theory(SBMFT) supplemented with Gutzwiller projection provides an exceedingly accurate description for the ground state of the spin-$\frac{1}{2}$ triangular lattice antiferromagnetic Heisenberg model(spin-$\frac{1}{2}$ TLHAF). However, we find the SBMFT fails even qualitatively in the description of the dynamical behavior of the system. In particular, the SBMFT fails to predict the Goldstone mode in the magnetic ordered phase. We show that the coherent peak in the two-spinon continuum in the presence of spinon condensate should not be interpreted as a magnon mode. The SBMFT also predicts incorrectly a gapless longitudinal spin fluctuation mode in the magnetic ordered phase. We show that these failures are related to the following facts:  (1)Spinon condensation fails to provide a consistent description of the order parameter manifold of the 120 degree ordered phase. (2)There lacks in the SBMFT the coupling between the uncondensed spinon  and the spinon condensate, which breaks both the spin rotational and the translational symmetry. (3)There lacks in the SBMFT the rigidity that is related to the no double occupancy constraint on the spinon system. We show that such failures of the SBMFT is neither restricted to the spin-$\frac{1}{2}$ TLHAF nor to the magnetic ordered phase. We proposed a generalized SBMFT to resolve the first two issues and a new formalism to address the third issue.
\end{abstract}

\maketitle

\section{Introduction}
The Schwinger Boson mean field theory(SBMFT) has been widely used in the study of quantum magnet systems\cite{Arovas,Auerbach,Liang,Read,Sachdev,Chen,Chandra,Coleman,Trumper1993,Gazza,Manuel,Trumper1997,Manuel1998,Wang,Tay,Tao,Mezio,Mezio2012,Messio,Zheng,Ghioldi,Bauer,Gonzalez,Ghioldi2018,Zhang}. As compared to its Fermionic cousin, the SBMFT has the advantage that it can describe the paramagnetic and the magnetic ordered phase on an equal footing. This becomes especially important when we are concerned with a quantum spin liquid in proximity to a magnetic ordered phase\cite{Tao}. The spin-$\frac{1}{2}$ antiferromagnetic Heisenberg model on the square lattice(spin-$\frac{1}{2}$ SLHAF) is a perfect example to the illustrate the power of the SBMFT. Gutzwiller projection of the mean field ground state of the SBMFT is found to be an extremely accurate description of the true ground state of the spin-$\frac{1}{2}$ SLHAF\cite{Liang,Chen}.  As will be shown in this paper, the same is true for the spin-$\frac{1}{2}$ antiferromagnetic Heisenberg model on the triangular lattice(spin-$\frac{1}{2}$ TLHAF). 

The success of the SBMFT in the description of the ground state property encourages people to use it to understand the dynamical behavior of quantum magnet system\cite{Arovas,Chandra,Coleman,Mezio,Mezio2012,Ghioldi,Gonzalez,Ghioldi2018,Zhang}. In recent years, it is found that many quantum magnet systems exhibit novel excitation behavior at high energy, even though they host conventional symmetry breaking order in the ground state and that the low energy excitation of the system is well described by the linear spin wave theory(LSWT) with the spin velocity properly renormalized. For example, roton-like minimum in the magnon dispersion is observed in both the spin-$\frac{1}{2}$ SLHAF system Cu(DCO$_2$)$_2$$\cdot$4D$_2$O (CFTD)\cite{Piazza} and the spin-$\frac{1}{2}$ TLHAF system Ba$_3$CoSb$_2$O$_9$\cite{Shirata,Susuki,Zhou,Ma,Saya,Kamiya}. Such roton-like minimum is found to be accompanied by intense spectral continuum at higher energy. Such spectral anomalies are hard to understand in the spin wave picture\cite{Jolicoeur,Zhitomirsky,ZhengW,Mourigal,Pollmann}, but may suggest the existence of fractionalized spin excitation\cite{Ho,Becca2019,Tao20201}.

However, people find that the SBMFT runs into serious problems when it is used to predict the dynamical  behavior of the quantum magnet system\cite{Chandra,Coleman,Ghioldi2018,Zhang}. According to the SBMFT, the spin fluctuation spectrum of the system is composed of the continuum of two spinon excitation. In the magnetic ordered phase, the lower edge of the two-spinon continuum becomes a coherent peak as one spinon condenses. Such a coherent peak has been interpreted as a magnon mode by many people. Indeed, in the case of the spin-$\frac{1}{2}$ SLHAF, the dispersion of such a peak has exactly the same form as the magnon dispersion predicted by the LSWT. However, this is just a coincidence\cite{Chandra,Coleman}. For the spin-$\frac{1}{2}$ TLHAF, the dispersion of such a peak is totally different from the magnon dispersion predicted by the LSWT, even in the large-S limit\cite{Ghioldi2018,Zhang}. More specifically, the SBMFT predicts an isotropic spin wave velocity in the long wave length limit, while we expect the magnon mode with dominate in-plane polarization character to propagate with a different velocity than the magnon mode with a dominate out-of-plane polarization character in the 120 degree ordered phase. In addition, the SBMFT predicts incorrectly that there are two, rather than three branches of linearly dispersing modes around the $\Gamma$ point and predicts a spurious quadratic mode around the $\mathbf{K}$ and $\mathbf{K}'$ point with a finite gap (which approaches zero in the large S limit). As we will see in this work, the SBMFT also predicts incorrectly a gapless longitudinal spin fluctuation mode in the magnetic ordered phase.

On a more conceptual level, the failure of the SBMFT in the magnetic ordered phase can be seen from its prediction of free spinon excitation in the low energy limit, which is expected to be confined in a magnetic ordered background. This is very different from the situation in the Fermionic RVB theory\cite{Ho,Tao20201}, in which symmetry breaking order parameter will gap out the Fermionic spinon in the low energy limit, leaving the Goldstone mode as the only spectral feature below the spinon gap. In fact, the Fermionic RVB theory can not only reproduce the Goldstone mode in the long wave length limit for the spin-$\frac{1}{2}$ TLHAF, but can also provide a qualitative picture for the observed spectral anomalies around the $\mathbf{M}$ point, if we assume a $\pi$-flux RVB structure in the ground state of the system\cite{Tao20201}. 

However, the Bosonic and the Fermionic RVB theory are actually closely related with each other. In particular, it can be shown that the short ranged RVB state on planar graph of both type can be identified with each other\cite{Anderson,Yunoki}. In this work, we show that the SBMFT provides an exceptionally accurate description of the ground state of the spin-$\frac{1}{2}$ TLHAF on finite system(or in the paramagnetic phase), when it is supplemented with the Gutzwiller projection onto the singly occupied subspace. More importantly, we find that the Bosonic RVB state so constructed can be continuously deformed into the short-ranged RVB state proposed first by Anderson\cite{Anderson}, which again can be deformed into the Dirac spin liquid state with the $\pi$-flux RVB structure proposed by Yunoki and Sorella\cite{Yunoki}. Thus the question is why the SBMFT describes the ground state of the spin-$\frac{1}{2}$ TLHAF so well but fails so badly when it is used to predict the dynamical behavior of the system?

Here we show that the failure of the SBMFT in the description of the dynamical behavior of the spin-$\frac{1}{2}$ TLHAF is related to the following three facts: (1)Spinon condensation can not describe consistently the order parameter manifold of the 120 degree ordered phase of the spin-$\frac{1}{2}$ TLHAF. (2)There lacks in the SBMFT the coupling between the uncondensed spinon and the spinon condensate, which breaks both the spin rotational and the translational symmetry of the system. (3)There lacks in the SBMFT the rigidity that is related to the no double occupancy constraint on the spinon system. We show that the failure of the SBMFT in predicting the dynamical behavior is neither restricted to the spin-$\frac{1}{2}$ TLHAF nor to the magnetic ordered phase. We have proposed a generalized SBMFT to resolve the first two issues and a new formalism to address the third issue. We show that the magnetic order will gap out the two-spinon continuum in the magnetic ordered phase, leaving the magnon, or, the collective fluctuation of the spinon system as a whole, as the only spectral feature in the low energy limit.

The paper is organized as follows. In the next section, we introduce the Bosonic RVB theory for the spin-$\frac{1}{2}$ TLHAF and present the result on the variational ground state at both the mean field level and that after the Gutzwiller projection. We show how accurate is the variational ground state generated from such a construction and how such an RVB state can be deformed to the short-ranged RVB state proposed by Anderson and the Dirac spin liquid state proposed by Yunoki and Sorella. In the third section, we discuss how and why the SBMFT fails to predict the correct dynamical behavior of the spin-$\frac{1}{2}$ TLHAF. In particular, we show that the spinon condensation can not describe consistently the order parameter manifold of the 120 degree ordered phase. In the fourth section, we propose a generalized SBMFT to resolve the first two issues listed in the last paragraph. We show that the magnetic order parameter will gap out the two-spinon continuum. We show that the lack of Boson rigidity in the mean field treatment leads inevitably to the false prediction of a gapless longitudinal spin fluctuation mode in the magnetic ordered phase and is responsible for the spurious quadratic mode around the  $\mathbf{K}$ and $\mathbf{K}'$ point. In the last section, we discuss the implication of our findings on the study of a general quantum magnet and propose a new formalism to address the issue of Boson rigidity.

\section{A Bosonic RVB theory of the spin-$\frac{1}{2}$ TLHAF}
The spin-$\frac{1}{2}$ TLHAF reads
\begin{equation}
H=\sum_{<i,j>}\mathbf{S}_{i}\cdot \mathbf{S}_{j},
\end{equation}
in which the sum is over nearest neighboring sites of the triangular lattice and we have set the exchange coupling as the unit of energy. In the Bosonic RVB formalism, we represent the spin in terms of Schwinger Boson operator as 
\begin{equation}
\mathbf{S}_{i}=\frac{1}{2}\sum_{\alpha,\beta}b^{\dagger}_{i,\alpha}\bm{\sigma}_{\alpha,\beta}b_{i,\beta}.
\end{equation}
Here $\alpha,\beta=\uparrow,\downarrow$ is the spin index of the Bosonic spinon operator, $\bm{\sigma}$ is the usual Pauli matrix. The spinon operator should satisfy the constraint of 
\begin{equation}
\sum_{\alpha}b^{\dagger}_{i,\alpha}b_{i,\alpha}=1
\end{equation} 
to be a faithful representation of the spin algebra. As we will see in this work, the no double occupancy constraint on the Bosonic spinon is crucial for a correct interpretation of the result obtained from the Bosonic RVB theory.

In terms of the Schwinger Boson representation, the model Hamiltonian of the spin-$\frac{1}{2}$ TLHAF can be written as
\begin{equation}
H=\sum_{<i,j>}(\hat{B}^{\dagger}_{i,j}\hat{B}_{i,j}-\hat{A}^{\dagger}_{i,j}\hat{A}_{i,j}),
\end{equation}
in which
\begin{eqnarray}
\hat{A}_{i,j}&=&\frac{1}{2}(b_{i,\uparrow}b_{j,\downarrow}-b_{i,\downarrow}b_{j,\uparrow})\nonumber\\
\hat{B}_{i,j}&=&\frac{1}{2}(b^{\dagger}_{i,\uparrow}b_{j,\uparrow}+b^{\dagger}_{i,\downarrow}b_{j,\downarrow}),
\end{eqnarray}
which suggests the mean field decoupling of $H$ in terms of the RVB order parameter $B_{i,j}=\langle\hat{B}_{i,j}\rangle$ and $A_{i,j}=\langle\hat{A}_{i,j}\rangle$. It is generally believed that both types of RVB order parameters are important for quantum magnet with a non-collinear magnetic ordering pattern. The mean field Hamiltonian takes the form of
\begin{eqnarray}
H_{MF}&=&\sum_{<i,j>}(B_{i,j}\hat{B}^{\dagger}_{i,j}-A_{i,j}\hat{A}^{\dagger}_{i,j}+h.c.)\nonumber\\
&+&\lambda\sum_{i,\alpha}b^{\dagger}_{i,\alpha}b_{i,\alpha},
\end{eqnarray}
in which $\lambda$ is a Lagrangian multiplier introduced to enforce the Boson number constraint at the mean field level. 

\begin{figure}
\includegraphics[width=9cm]{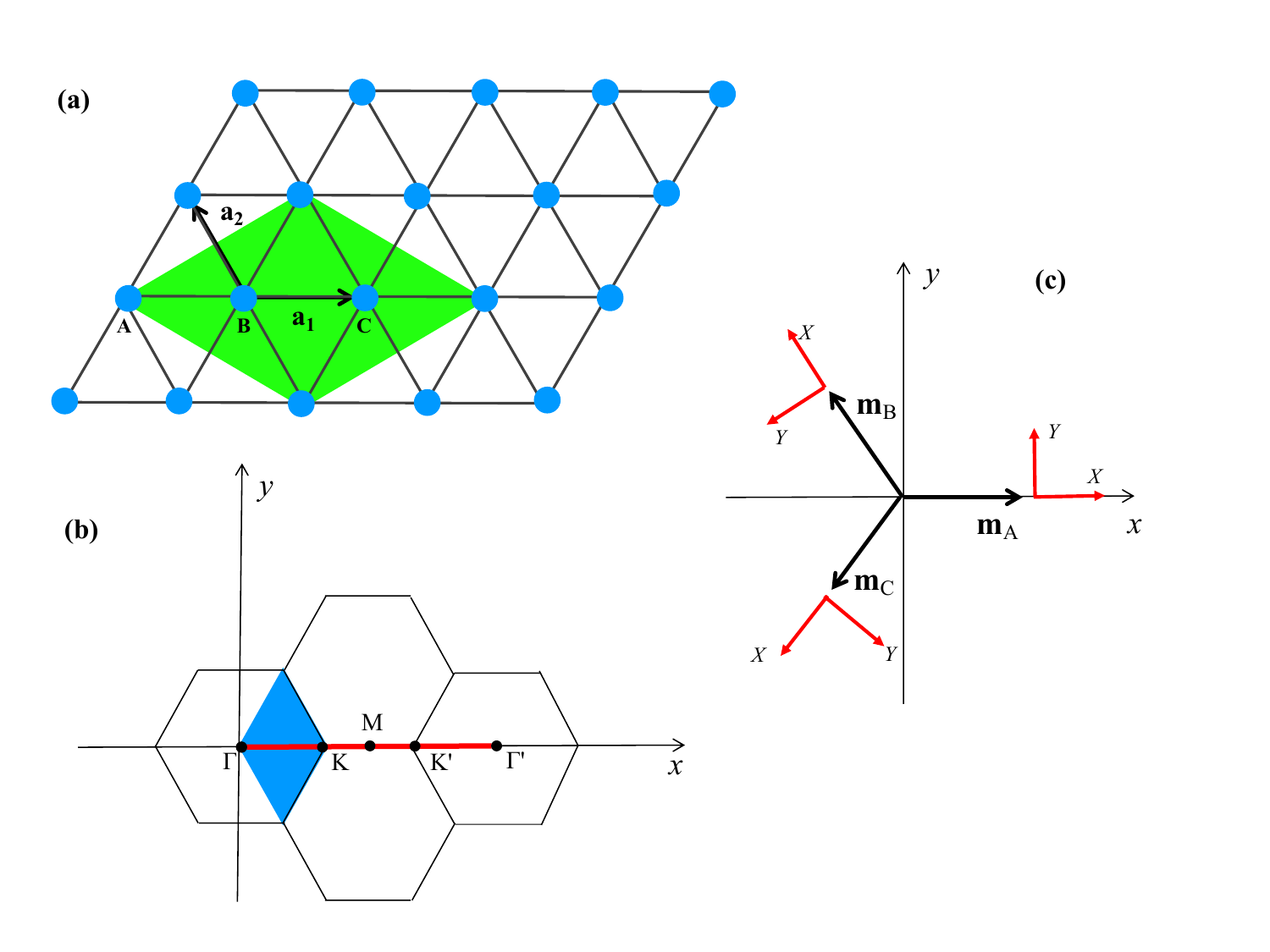}
\caption{Illustration of the magnetic unit cell(green area in (a)) and the magnetic Brillouin zone(blue area in (b)) in the 120 degree ordered phase of the spin-$\frac{1}{2}$ TLHAF. The magnetic unit cell contains three inequivalent sites A, B and C, whose ordered moment is shown in (c). Here $\mathbf{a}_{1}$ and $\mathbf{a}_{2}$ are the two basis vector of the lattice unit cell. The red vectors with the label $X$ and $Y$ in (c) are the unit vectors of the local coordinate system for the three sublattices. In the following discussion we will focus on the spin dynamics on the high symmetry path $\bm{\Gamma}-\mathbf{K}-\mathbf{K}'-\bm{\Gamma}'$ shown as the red line segment in (b). The $\mathbf{M}$ point is where the spin dynamics of the spin-$\frac{1}{2}$ TLHAF the most anomalous.}
\end{figure}

An ansatz for the RVB parameter $B_{i,j}$ and $A_{i,j}$ can be easily obtained by looking at the semiclassical limit of the model, in which we can take the spin operator $\mathbf{S}_{i}$ as classical vector and the boson operator $b_{i,\alpha}$ as conventional complex number. In the 120 degree ordered phase of the spin-$\frac{1}{2}$ TLHAF(as is illustrated in Fig.1), we get the following values of these complex numbers on the three sublattices of the system
\begin{eqnarray}
\left(\begin{array}{c}b_{i,\uparrow} \\b_{i,\downarrow}\end{array}\right)_{i\in A}&=&c_{i,A}\left(\begin{array}{c}\ 1 \ \\ \ 1\ \end{array}\right)\nonumber\\
\left(\begin{array}{c}b_{i,\uparrow} \\b_{i,\downarrow}\end{array}\right)_{i\in B}&=&c_{i,B}\left(\begin{array}{c}e^{-i\frac{\pi}{3}} \\e^{i\frac{\pi}{3}}\end{array}\right)\nonumber\\
\left(\begin{array}{c}b_{i,\uparrow} \\b_{i,\downarrow}\end{array}\right)_{i\in C}&=&c_{i,C}\left(\begin{array}{c}e^{-i\frac{2\pi}{3}} \\e^{i\frac{2\pi}{3}}\end{array}\right),
\end{eqnarray}
in which $c_{i,\mu}=\sqrt{m} e^{i\phi_{i,\mu}}$ for $\mu=A,B$ and $C$. Here $m$ denotes the length of the ordered magnetic moment. We note that the Schwinger Boson representation of the spin operator has an intrinsic $U(1)$ gauge redundancy as a result of the no double occupancy constraint on the spinon operator. $\phi_{i,\mu}$ is the corresponding gauge phase. The freedom to choose such a gauge phase can be exploited to simplify the form of the mean field ansatz.  In the following, we choose $\phi_{i,A}=\phi_{i,C}=\frac{\pi}{4}$ and $\phi_{i,B}=\frac{5\pi}{4}$ to make the RVB order parameter real and translational invariant.

Inserting the value of $b_{i,\alpha}$ we got from the semiclassical analysis in the definition of $A_{i,j}$ and $B_{i,j}$, we arrive at the following simple RVB mean field ansatz. 
\begin{eqnarray}
A_{i,i+\mathbf{a}_{1}}&=&A_{i,i+\mathbf{a}_{2}}=-A_{i,i+\mathbf{a}_{1}+\mathbf{a}_{2}}=A\nonumber\\
B_{i,i+\mathbf{a}_{1}}&=&B_{i,i+\mathbf{a}_{2}}=B_{i,i+\mathbf{a}_{1}+\mathbf{a}_{2}}=B,
\end{eqnarray}
in which $A$ and $B$ are both real numbers. We note that this ansatz has exactly the same form as that of the $A_{1}$ phase obtained from projective symmetry group analysis\cite{Wen,Wang,Zheng}. Furthermore, it reduces to the mean field ansatz proposed by Sachdev\cite{Sachdev} when we set $B=0$. We also note that while in the semiclassical limit $B=-\frac{1}{\sqrt{3}}A$, $B$ should in general be understood as an independent RVB parameter.

\begin{figure}
\includegraphics[width=8cm]{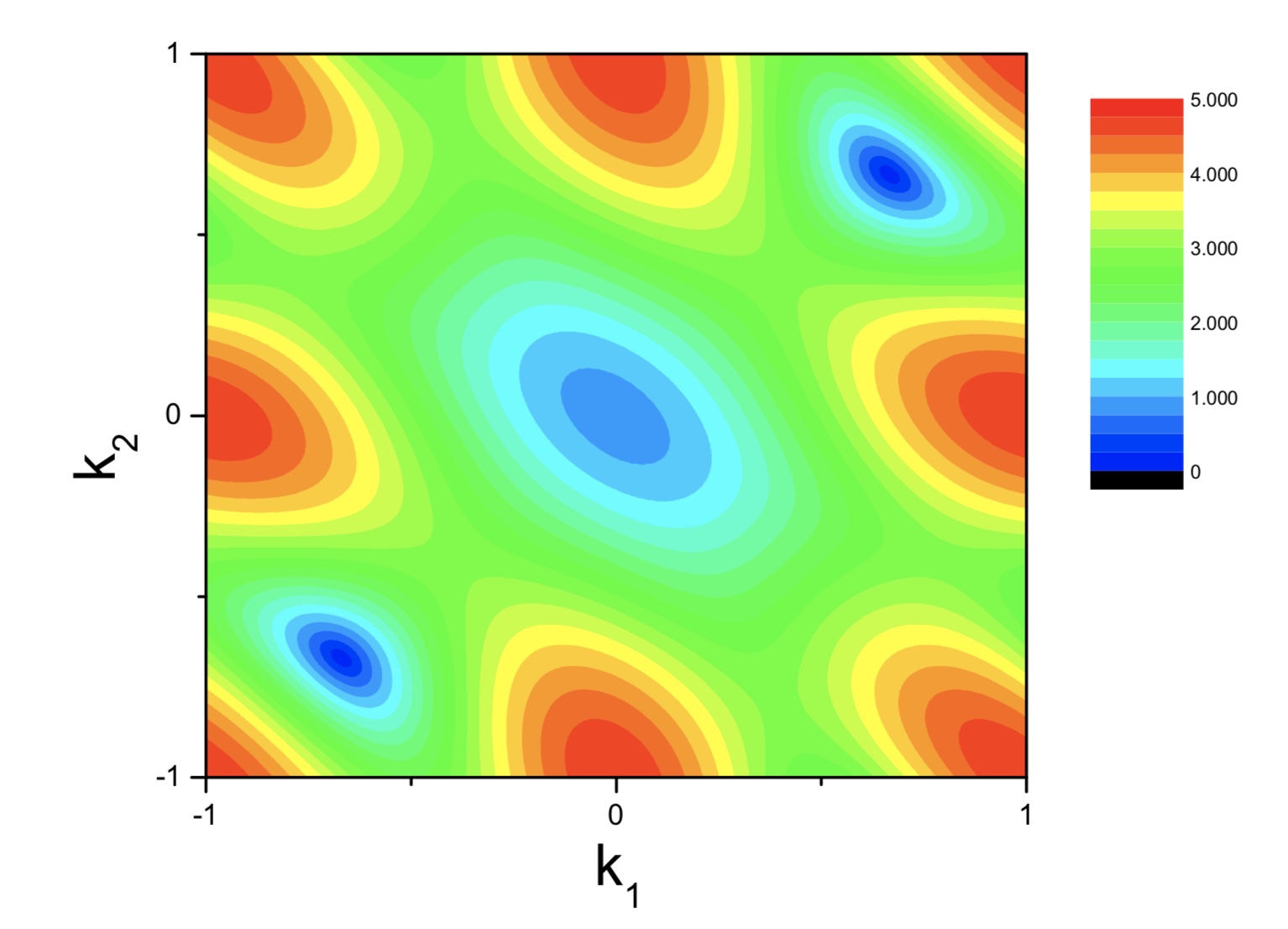}
\caption{The mean field spinon dispersion calculated at $B=-0.46A$, the optimum of mean field energy. Here we have set $A=1$. Here $k_{1}$ and $k_{2}$ are both plotted in unit of $\pi$.}
\end{figure}

In the momentum space, the mean field Hamiltonian of the spinon with the above mean field ansatz is given by
\begin{equation}
H_{MF}=\sum_{\mathbf{k}}\left(\begin{array}{cc}b^{\dagger}_{\mathbf{k},\uparrow} & b_{-\mathbf{k},\downarrow}\end{array}\right)\left(\begin{array}{cc}\xi_{\mathbf{k}} & \Delta^{*}_{\mathbf{k}} \\\Delta_{\mathbf{k}} & \xi_{\mathbf{k}}\end{array}\right)\left(\begin{array}{c}b_{\mathbf{k},\uparrow} \\b^{\dagger}_{-\mathbf{k},\downarrow}\end{array}\right)
\end{equation}
up to a constant. Here 
\begin{eqnarray}
\xi_{\mathbf{k}}&=&\lambda+B\gamma(\mathbf{k})\nonumber\\
\Delta_{\mathbf{k}}&=&A\eta(\mathbf{k}),
\end{eqnarray}
in which 
\begin{eqnarray}
\gamma(\mathbf{k})&=&2[\cos k_{1} +\cos k_{2} +\cos(k_{1}+k_{2})]\nonumber\\
\eta(\mathbf{k})&=&2i[\sin k_{1} +\sin k_{2} -\sin(k_{1}+k_{2})].
\end{eqnarray} 
Here $k_{1}=\mathbf{k}\cdot\mathbf{a}_{1}$ and $k_{2}=\mathbf{k}\cdot\mathbf{a}_{2}$ are the projection of $\mathbf{k}$ on the two basis vectors of the triangular lattice. Note that $\gamma(\mathbf{k})$ and $\eta(\mathbf{k})$ have opposite parity under inversion in the momentum space.
The eigen energy of $H_{MF}$ is given by 
\begin{equation}
\epsilon_{\mathbf{k}}=\sqrt{(\lambda+B\gamma(\mathbf{k}))^{2}-|A\eta_{\mathbf{k}}|^{2}}.
\end{equation}
It can be shown that for $B> -\frac{1}{\sqrt{3}}A$, the minimum of $\epsilon_{\mathbf{k}}$ is reached at $\mathbf{k}=\pm \mathbf{Q}=\pm(\frac{2\pi}{3},\frac{2\pi}{3})$(see Fig.2). For $B< -\frac{1}{\sqrt{3}}A$, the minimum of $\epsilon_{\mathbf{k}}$ moves to $\bm{\Gamma}=(0,0)$. Thus in the semiclassical limit, $\pm\mathbf{Q}$ is degenerate with $\bm{\Gamma}$ and both are gapless.

\begin{figure}
\includegraphics[width=7cm]{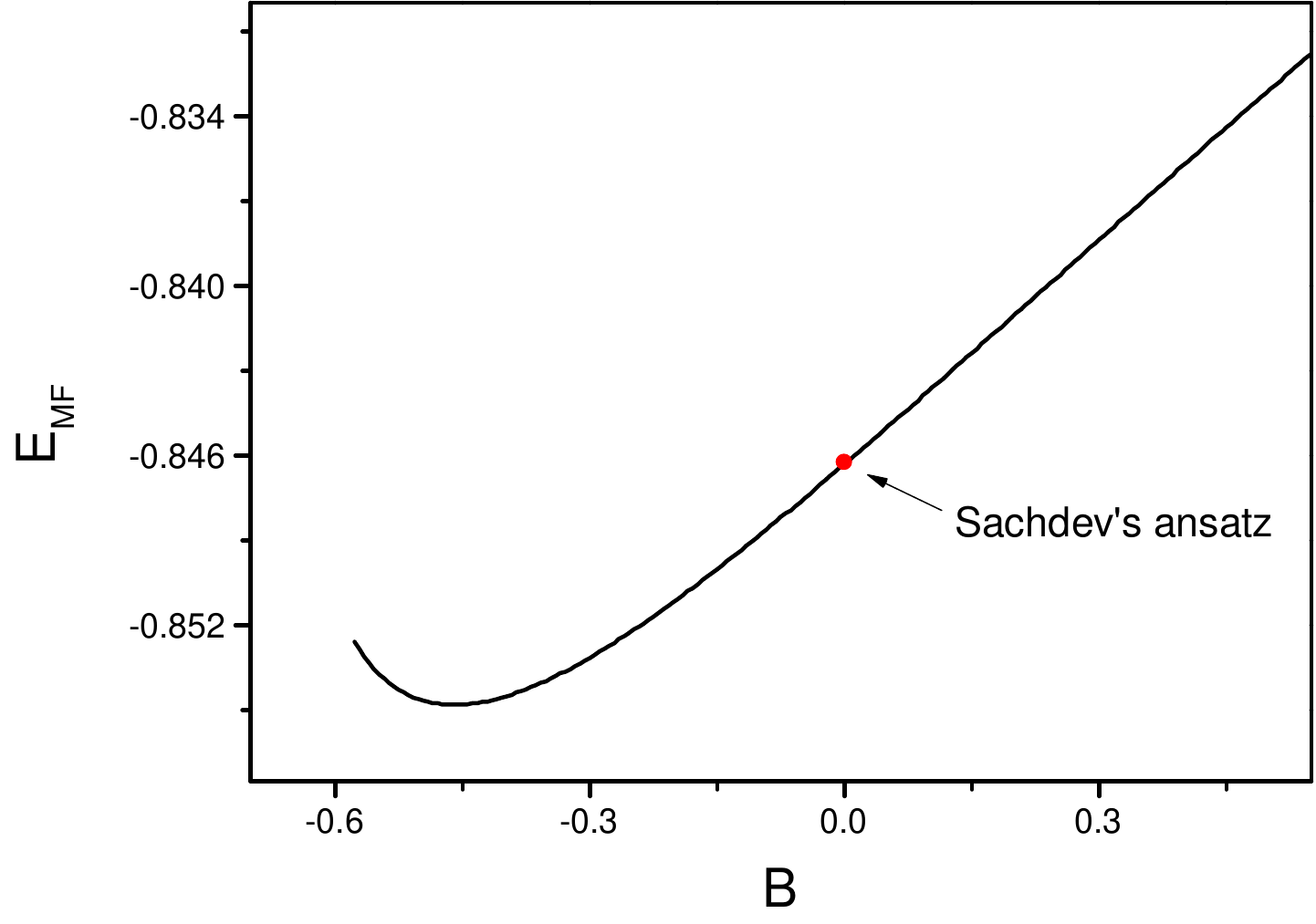}
\caption{The mean field ground state energy of the Bosonic RVB ansatz as a function of the RVB parameter $B$. Here we have set $A=1$. The red dot denotes the energy of the mean field ansatz proposed in Ref.[\onlinecite{Sachdev}].}
\end{figure}

The mean field ground state energy of the spin-$\frac{1}{2}$ TLHAF can be found from the Wick decomposition of the Heisenberg Hamiltonian in the mean field ground state and is given by
\begin{equation}
\langle \mathbf{S}_{i}\cdot\mathbf{S}_{i+\mathbf{a}_{1}}\rangle=\frac{3}{8}\left[  |\langle \hat{B}_{i,i+\mathbf{a}_{1}}\rangle|^{2}-|\langle\hat{A}_{i,i+\mathbf{a}_{1}}\rangle|^{2} \right]
\end{equation}
in which
\begin{eqnarray}
\langle \hat{A}_{i,i+\mathbf{a}_{1}}\rangle=\frac{1}{N}\sum_{\mathbf{k}}\frac{\Delta_{\mathbf{k}}^{*}}{\epsilon_{\mathbf{k}}}e^{i\mathbf{k}\cdot\mathbf{a}_{1}}\nonumber\\
\langle\hat{B}_{i,i+\mathbf{a}_{1}}\rangle=\frac{1}{N}\sum_{\mathbf{k}}\frac{\xi_{\mathbf{k}}}{\epsilon_{\mathbf{k}}}e^{i\mathbf{k}\cdot\mathbf{a}_{1}}.
\end{eqnarray}
The chemical potential is determined by the equation
\begin{equation}
1=\frac{1}{N}\sum_{\mathbf{k}}\frac{\xi_{\mathbf{k}}}{\epsilon_{\mathbf{k}}}.
\end{equation}
We find that the mean field solution satisfying the above equation is always gapless in the thermodynamic limit. The mean field ground state energy of the spin-$\frac{1}{2}$ TLHAF is found to be minimized around $B\approx-0.46A$ with a value of $E_{MF}\approx -0.855J/$site, which is slightly lower than that calculated from the Sachdev's ansatz with $B=0$(see Fig.3). 

The ground state energy calculated from the SBMFT is much lower than the true ground state energy of the spin-$\frac{1}{2}$ TLHAF(estimated to be about $-0.5458J/$site\cite{Capriotti}). However, we note that the SBMFT violates the local spin sum rule by a factor of $\frac{3}{2}$ in the mean field ground state\cite{Arovas}. More specifically, one find that in the SBMFT ground state
\begin{equation}
\langle \mathbf{S}_{i}\cdot\mathbf{S}_{i}\rangle=\frac{9}{8}=\frac{3}{2}\times\frac{3}{4}.
\end{equation}
If we rescale the mean field ground state energy by the same factor of $\frac{3}{2}$, it would become $-0.855J/$site$\times \frac{2}{3}\approx -0.57J$/site, which is very close to the best estimate\cite{Capriotti}. 

The accuracy of the Bosonic RVB theory in the description of the ground state of the spin-$\frac{1}{2}$ TLHAF can be seen more clearly from the Gutzwiller projected wave function derived from the mean field ground state. The Bosonic RVB state so constructed can be written in the form of
\begin{equation}
|RVB\rangle=P_{G}\left[ \sum_{i,j}a(\mathbf{r}_{i}-\mathbf{r}_{j})b^{\dagger}_{i,\uparrow}b^{\dagger}_{j,\downarrow}\right]^{N/2}|0\rangle,
\end{equation}
in which $N$ is the number of lattice site. $P_{G}$ is the Gutzwiller projection operator enforcing the no double occupancy constraint, $|0\rangle$ is the vacuum of the Schwinger Boson. The RVB amplitude $a(\mathbf{r}_{i}-\mathbf{r}_{j})$ is give by
\begin{equation}
a(\mathbf{r}_{i}-\mathbf{r}_{j})=\frac{1}{N}\sum_{\mathbf{k}}\frac{\Delta_{\mathbf{k}}}{\epsilon_{\mathbf{k}}+\xi_{\mathbf{k}}}e^{i\mathbf{k}\cdot(\mathbf{r}_{i}-\mathbf{r}_{j})}.
\end{equation}
To compute the ground state energy from such a variational state, we expand it in the Ising basis
\begin{equation}
|RVB\rangle=\sum_{c} \varphi(c) |c\rangle.
\end{equation}
Here $|c\rangle$ denotes a particular Ising spin configuration on the lattice. The wave function $\varphi(c)$ corresponding to such a configuration is given by
\begin{equation}
\varphi(c)=\mathrm{Per}\left(\begin{array}{cccc}. & . & . & . \\. & a(\mathbf{r}_{i}-\mathbf{r}_{j}) & . & . \\. & . & . & . \\. & . & . & .\end{array}\right).
\end{equation}
Here $\mathrm{Per}(M)$ denotes the permanent of a matrix $M$. $\mathbf{r}_{i}$ and $\mathbf{r}_{j}$ denotes the position of the $i$-th up spin and the $j$-th down spin in the spin configuration $|c \rangle$. 

 The variational energy of $|RVB\rangle$ can be expressed as 
\begin{equation}
E=\frac{\sum_{c}|\varphi(c)|^{2}\sum_{c'}\langle c |H |c'\rangle\frac{\varphi(c')}{\varphi(c)}}{\sum_{c}|\varphi(c)|^{2}},
\end{equation}
which can simulated by the standard variational Monte Carlo method\cite{Tay,Tao}.  On small clusters, the permanent of a matrix can be evaluated efficiently with the Ryser's algorithm. 

\begin{figure}
\includegraphics[width=8.5cm]{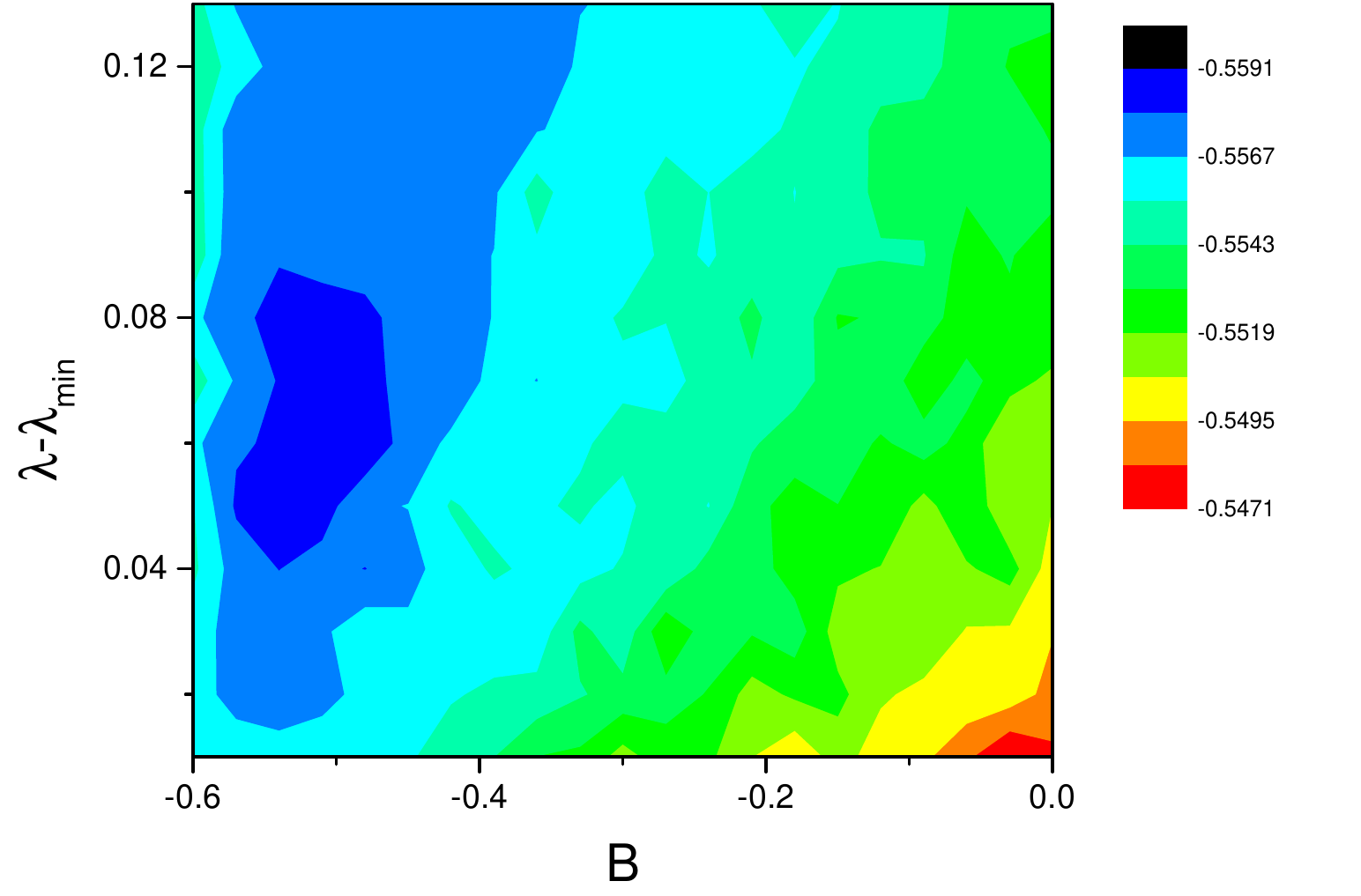}
\caption{The variational ground state energy of the Gutzwiller projected Bosonic RVB state as a function of the RVB parameter $B$ and the chemical potential $\lambda$. Here we have set the RVB parameter $A=1$. The VMC simulation is performed on a $6\times6$ cluster with periodic boundary condition in both directions. $2\times10^{6}$ samples are used in the evaluation of the variational energy. Note that the dependence of the variational energy with $B$ is rather weak.}
\end{figure}  

We have performed variational Monte Carlo simulation on such a Bosonic RVB state on a $6\times6$ cluster with periodic boundary condition. Here we treat both $B$ and $\lambda$ as variational parameter. The variational energy as a function of $B$ and $\lambda$ is shown in Fig.4. The minimum of the variational energy is found be realized around $B\approx -0.48A$, with a variational energy of $E_{min}\approx-0.559J/$site, which is extremely close to the exact result of $-0.5604J/$site obtained from exact diagonalization on the $6\times6$ cluster\cite{Capriotti}. The variational energy of the RVB state generated from the Sachdev's ansatz is slightly higher($-0.555J/$site). As a comparison, we note that the ground state energy on the same $6\times6$ cluster computed from the short-ranged RVB state, the Dirac spin liquid state, the Dirac spin liquid state with spin Jastrow factor correction are respectively $-0.521J/$site, $-0.528J/$site and $-0.537J/$site\cite{Yunoki}.  Beside the variational energy, we have also computed the overlap between the Bosonic RVB state and the exact ground state on the $6\times6$ cluster. We find that the overlap is as high as 0.978 for $B=-0.48A$. These results show collectively that the Bosonic RVB theory based on SBMFT provides a very accurate description on the ground state of the spin-$\frac{1}{2}$ TLHAF.

We note that the Bosonic RVB state discussed above can be deformed continuously into the short range RVB state proposed by P.W. Anderson in 1973\cite{Anderson}. This can be achieved by simply sending the chemical potential $\lambda$ to infinity, since 
\begin{equation}
a(\mathbf{r}_{i}-\mathbf{r}_{j})\rightarrow\frac{1}{2\lambda}\frac{1}{N}\sum_{\mathbf{k}}\Delta_{\mathbf{k}}e^{i\mathbf{k}\cdot(\mathbf{r}_{i}-\mathbf{r}_{j})}=\frac{A_{i,j}}{2\lambda}
\end{equation}
in the limit of $\lambda\rightarrow \infty$. Interestingly, such a short ranged RVB state can again be deformed continuously into the Dirac spin liquid state proposed by Yunoki and Sorella\cite{Yunoki}. Such a Fermionic RVB state is generated from Gutzwiller projection of the ground state of the following mean field Hamiltonian
\begin{equation}
H_{\pi-flux}=-\sum_{<i,j>,\alpha}\eta_{i,j}(f^{\dagger}_{j,\alpha}f_{i,\alpha}+h.c.),
\end{equation} 
in which $f_{i,\alpha}$ denotes a Ferminoic spinon operator with spin $\alpha$, $\eta_{i,j}=\pm1$ is the sign of the Fermion hopping integral between site $i$ and $j$, chosen in such a way that each elementary plaquette of the triangular lattice encloses a gauge flux of $\pi$(see Fig.5). In a recent work\cite{Li}, we find that such a $\pi$-flux structure in the Fermionic RVB theory is responsible for the observed spectral anomaly of the spin-$\frac{1}{2}$ TLHAF around the $\mathbf{M}$ point\cite{Tao20201}. The close relation between the Bosonic RVB state studied in this work with such a Fermionic RVB state implies that the Bosonic RVB theory may also provide a satisfactory explanation of the spectral anomalies of the spin-$\frac{1}{2}$ TLHAF around the $\mathbf{M}$ point.

\begin{figure}
\includegraphics[width=9cm]{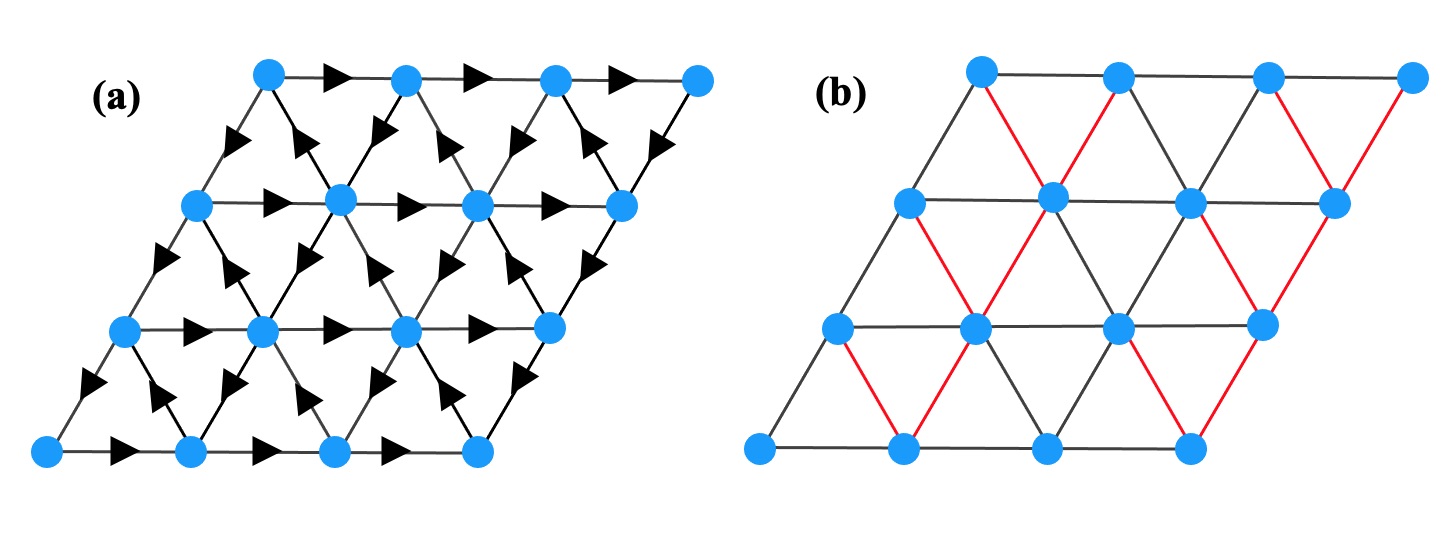}
\caption{Illustration of the structure of short range RVB state proposed by Anderson(a) and the Dirac spin liquid state proposed by Yunoki and Sorella(b) for the spin-$\frac{1}{2}$ TLHAF. For the short range RVB state, we plot the sign of $A_{i,j}$ with arrow. $A_{i,j}$ is positive in the direction of the arrow and is otherwise negative(note $A_{i,j}$ is antisymmetric with respect to the exchange of $i$ and $j$).  For the Dirac spin liquid state, we plot the sign of the hopping integral $\eta_{i,j}$ with color. $\eta_{i,j}$ is positive when the bond is plotted black and is otherwise negative(note that $\eta_{i,j}$ is symmetric with respect to the exchange of $i$ and $j$).}
\end{figure}

\begin{figure}
\includegraphics[width=7cm]{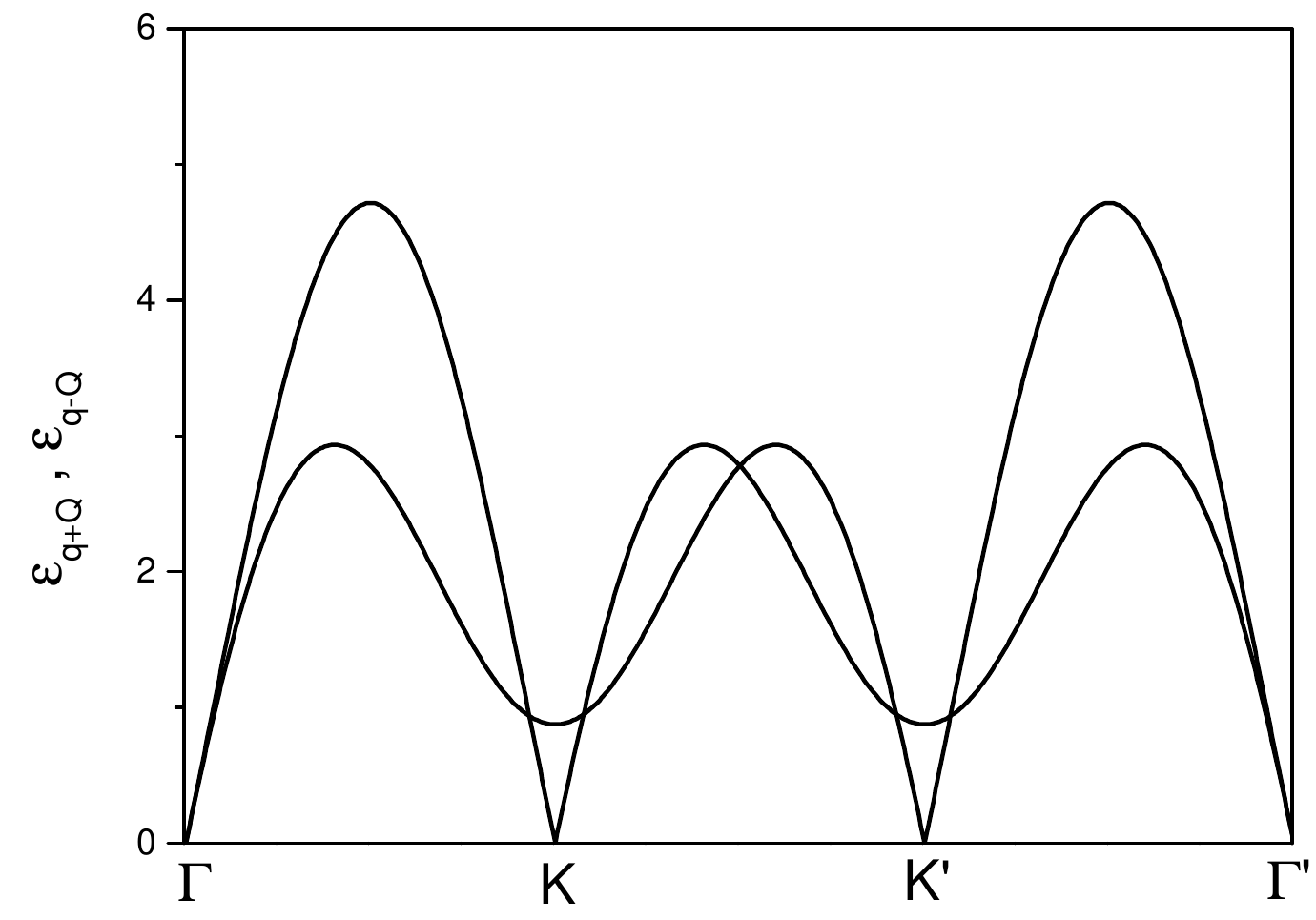}
\caption{The magnon dispersion predicted by the SBMFT along the high symmetry path $\bm{\Gamma}-\mathbf{K}-\mathbf{K'}-\bm{\Gamma}'$ in the Brillouin zone. Here we have set the RVB order parameter $A=1$ and choose $B=-0.46$. The coherent peak emerges at the edge of the two-spinon continuum when one spinon condenses at momentum $\mathbf{Q}$ or $-\mathbf{Q}$. The dispersion of the two coherent peak in the SBMFT are thus given by $\epsilon_{\mathbf{q+Q}}$ and $\epsilon_{\mathbf{q-Q}}$, in which $\epsilon_{\mathbf{k}}$ is the dispersion of a single spinon in the SBMFT.}
\end{figure}  

\section{Why the SBMFT fails to predict the correct magnon dispersion of the spin-$\frac{1}{2}$ TLHAF?}
The amazing success of the SBMFT in describing the ground state property of the spin-$\frac{1}{2}$ TLHAF seems to imply its success in the description of the dynamical behavior of the system. However, this is not the case. The spin fluctuation spectrum of the spin-$\frac{1}{2}$ TLHAF calculated from the SBMFT is composed of the continuum of two-spinon excitation. In the magnetic ordered phase, the lower edge of the two-spinon continuum becomes a coherent peak as one spinon condenses at momentum $\mathbf{Q}$ or $-\mathbf{Q}$. Such coherent peaks have been interpreted as the magnon mode of the spin-$\frac{1}{2}$ TLHAF by many people. In Fig.6 we plot the dispersion of these modes along the high symmetry path $\bm{\Gamma}-\mathbf{K}-\mathbf{K'}-\bm{\Gamma}'$ in the Brillouin zone. 

The prediction of the SBMFT has the following problems\cite{Ghioldi2018,Zhang}. (1)The SBMFT predicts that there are only two, rather than three branches of magnon mode, as is expected in the 120 degree ordered phase of the spin-$\frac{1}{2}$ TLHAF. (2)The SBMFT predicts incorrectly an isotropic spin wave velocity around the $\bm{\Gamma}$ point, while we expect that the magnon mode with dominate in-plane polarization character to propagate with a different velocity than that with a dominate out-of-plane polarization character in the 120 degree ordered phase. (3)The SBMFT predicts a spurious modes with a quadratic dispersion around the $\mathbf{K}$ and $\mathbf{K}'$ point, which becomes gapless in the semiclassical limit. (4)As we will show in the next section, the SBMFT also predicts incorrectly a gapless longitudinal spin fluctuation mode in the magnetic ordered phase. (5)The SBMFT predicts incorrectly the coexistence of spinon continuum and magnon mode in the low energy limit, while we expect that the spinon should be confined in the magnetic ordered phase. 

As we will see below, these problems are related to the following facts: (1)The spinon condensation fails to provide a consistent description of the order parameter manifold of the 120 degree ordered phase of the spin-$\frac{1}{2}$ TLHAF. (2)There lacks in the SBMFT the coupling between the uncondensed spinon and the spinon condensate, which breaks both the spin rotational and the translational symmetry. (3)There lacks in the SBMFT the rigidity  that is related to the no double occupancy constraint on the spinon system. Here we will focus on the first point and will leave the discussion of the other two points to the next section.

The description of the low energy physics of the magnetic ordered phase relies crucially on the proper description of the order parameter manifold, since the Goldstone mode is nothing but the global rotation of the system within such a space. In the SBMFT of the spin-$\frac{1}{2}$ TLHAF, spinon can condense at both $\mathbf{Q}$  and $-\mathbf{Q}$. Thus the spinon condensate is described by four independent  complex number $\bar{b}_{\pm\mathbf{Q},\sigma}=\langle b_{\pm\mathbf{Q},\sigma} \rangle$, in which $\sigma=\pm 1$ denotes the spin index of the spinon operator. The contribution of the spinon condensate to the Boson density and the spin density are given by
\begin{eqnarray}
\langle n_{i}\rangle_{c}&=&\frac{1}{N}\sum_{s,\sigma} \left[ \ |\bar{b}_{s\mathbf{Q},\sigma}|^{2}+ \bar{b}^{*}_{-s\mathbf{Q},\sigma} \bar{b}_{s\mathbf{Q},\sigma} e^{2is\mathbf{Q}\cdot \mathbf{R}_{i}} \ \right]  \nonumber\\
\langle \mathrm{S}^{z}_{i}\rangle_{c}&=&\frac{1}{2N}\sum_{s,\sigma}\sigma \left[ \ |\bar{b}_{s\mathbf{Q},\sigma}|^{2}+ \bar{b}^{*}_{-s\mathbf{Q},\sigma} \bar{b}_{s\mathbf{Q},\sigma} e^{2is\mathbf{Q}\cdot \mathbf{R}_{i}} \ \right]  \nonumber\\
\langle \mathrm{S}^{+}_{i}\rangle_{c}&=&\frac{1}{N}\sum_{s} \left[ \ \bar{b}^{*}_{s\mathbf{Q},\uparrow}\bar{b}_{s\mathbf{Q},\downarrow}+\bar{b}^{*}_{-s\mathbf{Q},\uparrow} \bar{b}_{s\mathbf{Q},\downarrow} e^{2is\mathbf{Q}\cdot \mathbf{R}_{i}} \ \right],  \nonumber\\
\end{eqnarray}
in which $s=\pm1$. Thus, the spinon condensate can not only generate the 120 degree magnetic order, but can also result in a modulation in the spinon density with wave vector $\pm\mathbf{Q}$ and a ferromagnetic order, neither is allowed in the 120 degree ordered phase of the spin-$\frac{1}{2}$ TLHAF. We thus have the following constraint on the spinon condensate
\begin{eqnarray}
\sum_{s,\sigma}  \ |\bar{b}_{s\mathbf{Q},\sigma}|^{2}&=&N_{c}\nonumber\\
\sum_{s,\sigma}  \sigma |\bar{b}_{s\mathbf{Q},\sigma}|^{2}&=&0\nonumber\\
\sum_{\sigma}\bar{b}^{*}_{-\mathbf{Q},\sigma} \bar{b}_{\mathbf{Q},\sigma}&=&0\nonumber\\
\sum_{s}\ \bar{b}^{*}_{s\mathbf{Q},\uparrow}\bar{b}_{s\mathbf{Q},\downarrow}&=&0,
\end{eqnarray}  
in which $N_{c}=N\langle n_{i}\rangle_{c}$ measures the overall strength of the spinon condensation. These constraints constitute 6 real equations. Thus we are left with only two independent real variables, which is insufficient to describe the order parameter manifold of the 120 degree ordered phase. More specifically, one need two real variables to describe the orientation of the 120 degree ordering plane and one additional real variable to describe the direction of the ordered moment within the ordering plane in such a phase.

Before moving on, let us solve the above constraint to see what kind of magnetic ordering pattern can be consistently described by single spinon condensation. To solve the first two constraints, we rewrite $b_{s\mathbf{Q},\sigma}$ as
\begin{eqnarray}
\bar{b}_{\mathbf{Q},\uparrow}&=&\sqrt{\frac{N_{c}}{2}}\cos \theta_{\uparrow} e^{i\phi_{\mathbf{Q}\uparrow}}\nonumber\\
\bar{b}_{-\mathbf{Q},\uparrow}&=&\sqrt{\frac{N_{c}}{2}}\sin \theta_{\uparrow} e^{i\phi_{-\mathbf{Q}\uparrow}}\nonumber\\
\bar{b}_{\mathbf{Q},\downarrow}&=&\sqrt{\frac{N_{c}}{2}}\cos \theta_{\downarrow} e^{i\phi_{\mathbf{Q}\downarrow}}\nonumber\\
\bar{b}_{-\mathbf{Q},\downarrow}&=&\sqrt{\frac{N_{c}}{2}}\sin \theta_{\downarrow} e^{i\phi_{-\mathbf{Q}\downarrow}}.
\end{eqnarray}
Here $\theta_{\uparrow},\theta_{\downarrow}\in [0,\frac{\pi}{2}]$. Inserting these expressions in Eq.25, we arrive at the following equations
\begin{eqnarray}
\cos \theta_{\uparrow} \sin \theta_{\uparrow} e^{i\phi}+\cos \theta_{\downarrow} \sin \theta_{\downarrow}&=&0\nonumber\\
\cos \theta_{\uparrow} \cos \theta_{\downarrow} e^{i\phi}+\sin \theta_{\uparrow} \sin \theta_{\downarrow}&=&0,
\end{eqnarray}
in which $\phi=\phi_{\mathbf{Q}\uparrow}-\phi_{-\mathbf{Q}\uparrow}+\phi_{-\mathbf{Q}\downarrow}-\phi_{\mathbf{Q}\downarrow}$. As we will show below, there are only three sets of solution to the above equations.

The first solution is 
\begin{eqnarray}
\theta_{\uparrow}=0,\ \  \theta_{\downarrow}=\frac{\pi}{2}, \  \ \phi \ \ \mathrm{arbitrary}.
\end{eqnarray}
The correspnding spinon condensate is given by
\begin{eqnarray}
\bar{b}_{\mathbf{Q},\uparrow}&=&\sqrt{\frac{N_{c}}{2}}e^{i\phi_{\mathbf{Q}\uparrow}}\nonumber\\
\bar{b}_{-\mathbf{Q},\uparrow}&=&0\nonumber\\
\bar{b}_{\mathbf{Q},\downarrow}&=&0\nonumber\\
\bar{b}_{-\mathbf{Q},\downarrow}&=&\sqrt{\frac{N_{c}}{2}}e^{i\phi_{-\mathbf{Q}\downarrow}}.
\end{eqnarray}
This solution describes a 120 degree order in the $x-y$ plane. More specifically, we have
\begin{eqnarray}
\langle \mathrm{S}^{z}_{i}\rangle&=&0\nonumber\\
\langle \mathrm{S}^{+}_{i}\rangle&=&\frac{N_{c}}{2N}e^{-i\left[ 2\mathbf{Q}\cdot \mathbf{R}_{i}+(\phi_{\mathbf{Q}\uparrow}-\phi_{-\mathbf{Q}\downarrow}) \right]}
\end{eqnarray}

The second solution is 
\begin{eqnarray}
\theta_{\uparrow}=\frac{\pi}{2},\ \  \theta_{\downarrow}=0, \  \ \phi \ \ \mathrm{arbitrary}.
\end{eqnarray}
The corresponding spinon condensate is given by
\begin{eqnarray}
\bar{b}_{\mathbf{Q},\uparrow}&=&0\nonumber\\
\bar{b}_{-\mathbf{Q},\uparrow}&=&\sqrt{\frac{N_{c}}{2}}e^{i\phi_{-\mathbf{Q}\uparrow}}\nonumber\\
\bar{b}_{\mathbf{Q},\downarrow}&=&\sqrt{\frac{N_{c}}{2}}e^{i\phi_{\mathbf{Q}\downarrow}}\nonumber\\
\bar{b}_{-\mathbf{Q},\downarrow}&=&0.
\end{eqnarray}
This solution also describes a 120 degree order of the spin in the $x-y$ plane. More specifically, we have
\begin{eqnarray}
\langle \mathrm{S}^{z}_{i}\rangle&=&0\nonumber\\
\langle \mathrm{S}^{+}_{i}\rangle&=&\frac{N_{c}}{2N}e^{i\left[ 2\mathbf{Q}\cdot \mathbf{R}_{i}+(\phi_{\mathbf{Q}\downarrow}-\phi_{-\mathbf{Q}\uparrow}) \right]}
\end{eqnarray}

The third solution is more nontrivial and is given by 
\begin{eqnarray}
\theta_{\uparrow}=\frac{\pi}{4},\ \ \theta_{\downarrow}=\frac{\pi}{4}, \ \ \phi=\pi.
\end{eqnarray}
The corresponding spinon condensate is given by
\begin{eqnarray}
\bar{b}_{\mathbf{Q},\uparrow}&=&\sqrt{\frac{N_{c}}{4}}e^{i\phi_{\mathbf{Q}\uparrow}}\nonumber\\
\bar{b}_{-\mathbf{Q},\uparrow}&=&\sqrt{\frac{N_{c}}{4}}e^{i\phi_{-\mathbf{Q}\uparrow}}\nonumber\\
\bar{b}_{\mathbf{Q},\downarrow}&=&\sqrt{\frac{N_{c}}{4}}e^{i\phi_{\mathbf{Q}\downarrow}}\nonumber\\
\bar{b}_{-\mathbf{Q},\downarrow}&=&\sqrt{\frac{N_{c}}{4}}e^{i\phi_{-\mathbf{Q}\downarrow}}.
\end{eqnarray}
with  $\phi_{\mathbf{Q}\uparrow}-\phi_{-\mathbf{Q}\uparrow}+\phi_{-\mathbf{Q}\downarrow}-\phi_{\mathbf{Q}\downarrow}=\phi=\pi$. This solution describes a 120 degree order with its ordering plane perpendicular to the $x-y$ plane. More specifically, we have
\begin{eqnarray}
\langle \mathrm{S}^{z}_{i}\rangle&=&\frac{N_{c}}{2N}\cos \Theta_{i} \nonumber\\
\langle \mathrm{S}^{x}_{i}\rangle&=&\frac{N_{c}}{2N}\sin \Theta_{i}\cos \Phi\nonumber\\
\langle \mathrm{S}^{y}_{i}\rangle&=&\frac{N_{c}}{2N}\sin \Theta_{i}\sin \Phi
\end{eqnarray}
in which $\Theta_{i}= 2\mathbf{Q}\cdot\mathbf{R}_{i}-\phi_{-\mathbf{Q}\uparrow}$, $\Phi=\phi_{\mathbf{Q}\downarrow}+\frac{\pi}{2}$. Here we have choose a gauge in which $\phi_{\mathbf{Q}\uparrow}=0$.

\section{A genralized SBMFT theory of the spin-$\frac{1}{2}$ TLHAF}
In the last section, we show that spinon condensation alone is insufficient to describe the order parameter manifold of the 120 degree ordered phase. At the same time, since the uncondensed spinon is decoupled from the spinon condensate in the SBMFT, it is actually propagating in a spin rotational and translational invariant environment. This explains why SBMFT predicts an isotropic magnon velocity and an incorrect number 2 of magnon branch in the 120 degree ordered phase. In the Fermionic RVB theory\cite{Ho,Tao20201}, these problems are solved by introducing the magnetic order parameter into the mean field Hamiltonian, which will gap out the spinon excitation, endow the magnon mode with an anisotropic velocity, and induce the folding needed to produce the correct number of magnon branch in the magnetic ordered phase. It is interesting to see if the magnetic order can play a similar role in the SBMFT description.

With such a consideration in mind, we propose the following generalized Schwinger Boson mean field theory for the symmetry breaking phase of the spin-$\frac{1}{2}$ TLHAF. Just as what we would do in the Fermionic RVB theory, we introduce the ordered moment $\langle\mathbf{S}_{i}\rangle$ as an additional order parameter besides the RVB order parameter $A_{i,j}$ and $B_{i,j}$. The mean field Hamiltonian in the generalized SBMFT theory reads
\begin{eqnarray}
H_{MF}&=&\sum_{<i,j>}(B_{i,j}\hat{B}^{\dagger}_{i,j}-A_{i,j}\hat{A}^{\dagger}_{i,j}+h.c.)\nonumber\\
&+&\lambda\sum_{i,\alpha}b^{\dagger}_{i,\alpha}b_{i,\alpha}-\mathrm{J_{eff}}\sum_{<i,j>}\langle\mathbf{S}_{i}\rangle\cdot\mathbf{S}_{j},
\end{eqnarray}
in which $\mathrm{J_{eff}}$ should be understood as a phenomenological parameter to avoid possible double counting on the effect of the Heisenberg exchange coupling. In principle, both the RVB order parameter $A_{i,j}$ and $B_{i,j}$ and the ordered moment $\langle\mathbf{S}_{i}\rangle$ should be understood as variational parameters to be determined by optimizing the variational ground state energy. Here we will be satisfied with a phenomenological treatment in which the RVB parameter $A_{i,j}$ and $B_{i,j}$ are fixed at their optimized values when  $\langle\mathbf{S}_{i}\rangle=0$.

In the 120 degree ordered phase, it is more convenient to describe the system in terms of the magnetic unit cell rather than the lattice unit cell. The mean field Hamiltonian of the spin-$\frac{1}{2}$ TLHAF has the form of
\begin{equation}
H_{MF}=\sum_{\mathbf{k}}\Psi^{\dagger}_{\mathbf{k}} \mathbf{H}_{\mathbf{k}}\Psi_{\mathbf{k}},
\end{equation}
in which 
\begin{equation}
\Psi_{\mathbf{k}}=\left(\begin{array}{c} \bm{\phi}_{\mathbf{k}} \\ \bm{\phi}^{\dagger}_{\mathbf{-k}}\end{array}\right).
\end{equation}
Here $\bm{\phi}_{\mathbf{k}}$ is given by
\begin{equation}
\bm{\phi}_{\mathbf{k}}=\left(\begin{array}{c}b_{\mathbf{k},A,\uparrow} \\b_{\mathbf{k},B,\uparrow} \\b_{\mathbf{k},C,\uparrow} \\b_{\mathbf{k},A,\downarrow} \\b_{\mathbf{k},B,\downarrow} \\b_{\mathbf{k},C,\downarrow}\end{array}\right).
\end{equation}
$\mathbf{H}_{\mathbf{k}}$ is a $12\times12$ matrix and is given by
\begin{equation}
\mathbf{H}_{\mathbf{k}}=\left(\begin{array}{cccc}M_{1} & M_{2} & 0 & M_{3} \\M^{*}_{2} & M_{1} & -M_{3} & 0 \\0 & M_{3} & M_{1} & M^{*}_{2} \\-M_{3} & 0 & M_{2} & M_{1}\end{array}\right),
\end{equation}
in which $M_{1}$, $M_{2}$ and $M_{3}$ are given by
\begin{eqnarray}
M_{1}&=&\lambda+B\left(\begin{array}{ccc} 0 \ \ & f^{*}_{\mathbf{k}}  \ \ & g^{*}_{\mathbf{k}} \\ f_{\mathbf{k}} \ \ & 0  \ \ & f^{*}_{\mathbf{k}} \\ g_{\mathbf{k}}  \ \ &f_{\mathbf{k}}  \ \ & 0\end{array}\right)\nonumber\\
M_{2}&=&D\left(\begin{array}{ccc}1 & 0 & 0 \\ 0 & e^{-i\frac{2\pi}{3}} & 0 \\ 0 & 0 & e^{-i\frac{4\pi}{3}}\end{array}\right) \nonumber\\
M_{3}&=&A\left(\begin{array}{ccc}0 & -f^{*}_{\mathbf{k}} & g^{*}_{\mathbf{k}} \\ f_{\mathbf{k}} & 0 & -f^{*}_{\mathbf{k}} \\ -g_{\mathbf{k}} & f_{\mathbf{k}} & 0\end{array}\right).
\end{eqnarray}
Here $f_{\mathbf{k}}=1+e^{ik_{1}}+e^{ik_{2}}$, $g_{\mathbf{k}}=e^{ik_{1}}+e^{ik_{2}}+e^{i(k_{1}+k_{2})}$, $D=-\frac{3\mathrm{J_{eff}m}}{2}$, with $m=|\langle \mathbf{S}_{i} \rangle|$. In the following, we will set $A=1$ as the unit of energy for the spinon Hamiltonian and set $B=-0.46$. In addition, we will treat $\mathrm{J_{eff}m}$ as a free parameter.

The mean field Hamiltonian can be diagonalized by the following para-unitary transformation
\begin{equation}
\Psi_{\mathbf{k}}=U_{\mathbf{k}}\Phi_{\mathbf{k}},
\end{equation}
in which the matrix $U_{\mathbf{k}}$ satisfy the equation
\begin{eqnarray}
U_{\mathbf{k}}\ \Sigma \ U^{\dagger}_{\mathbf{k}}&=&\Sigma \nonumber\\
U^{\dagger}_{\mathbf{k}}\ \mathbf{H}_{\mathbf{k}}\ U_{\mathbf{k}}&=&\Lambda_{\mathbf{k}}.
\end{eqnarray}
Here
\begin{equation}
\Sigma=\left(\begin{array}{cc}\mathbf{I} & 0 \\0 & -\mathbf{I}\end{array}\right),\Lambda_{\mathbf{k}}=\left(\begin{array}{cc}\bm{\epsilon}_{\mathbf{k}} & 0 \\0 & \bm{\epsilon}_{\mathbf{k}}\end{array}\right),
\end{equation}
in which $\mathbf{I}$ a $6\times6$ identity matrix, $\bm{\epsilon}_{\mathbf{k}}$ is a $6\times6$ diagonal matrix with diagonal element $\varepsilon^{n}_{\mathbf{k}}$ for $n=1,...,6$. The chemical potential $\lambda$ should be determined by the equation
\begin{eqnarray}
6&=&\frac{1}{2N}\sum_{\mathbf{k}}\langle \Psi^{\dagger}_{\mathbf{k}} \  \Psi_{\mathbf{k}}\rangle\nonumber\\
&=&\frac{1}{2N}\sum_{\mathbf{k}}\sum^{12}_{n=7} \left[ U^{\dagger}_{\mathbf{k}} U_{\mathbf{k}}\right]_{n,n}.
\end{eqnarray}

\begin{figure}
\includegraphics[width=7cm]{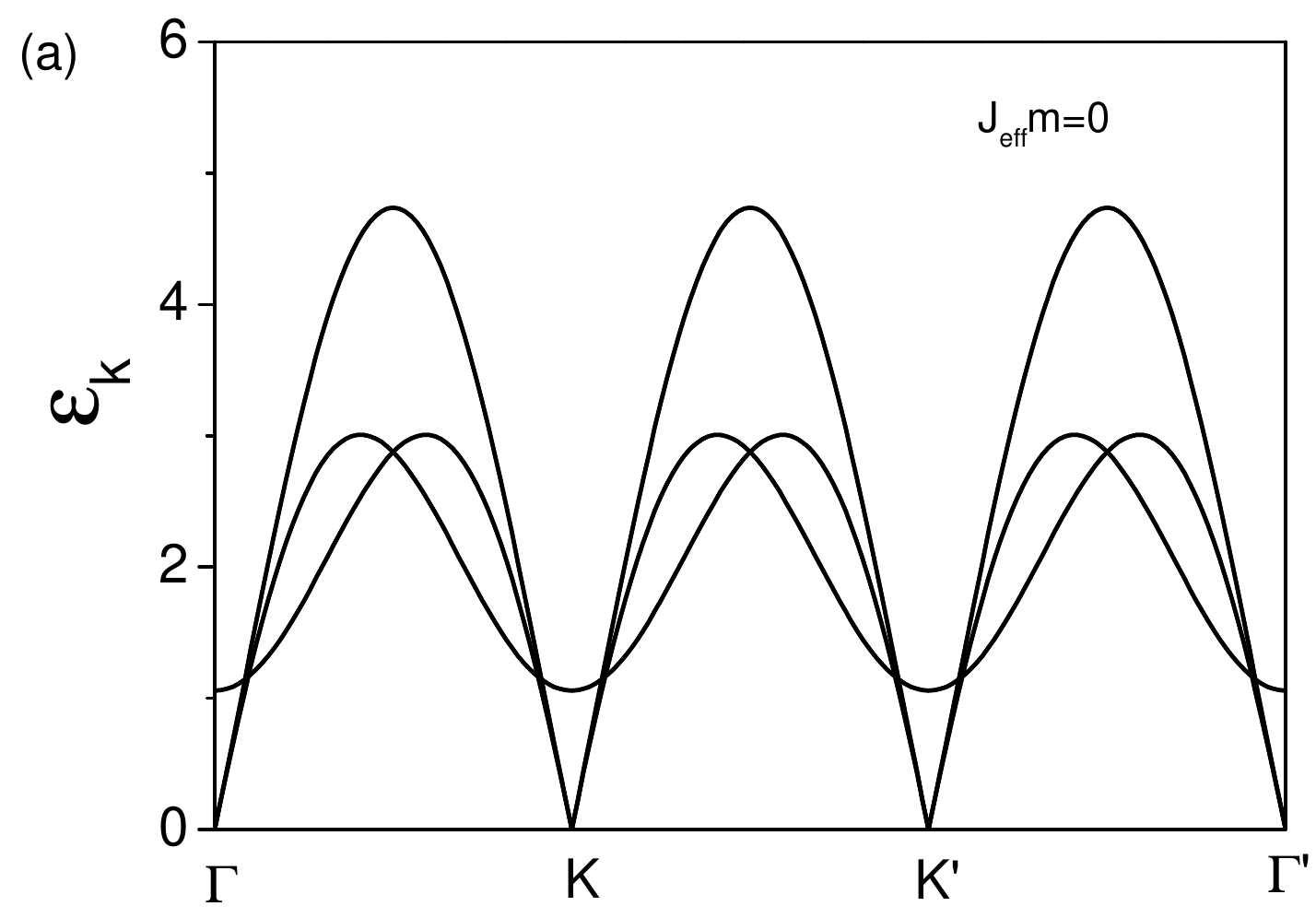}
\includegraphics[width=7cm]{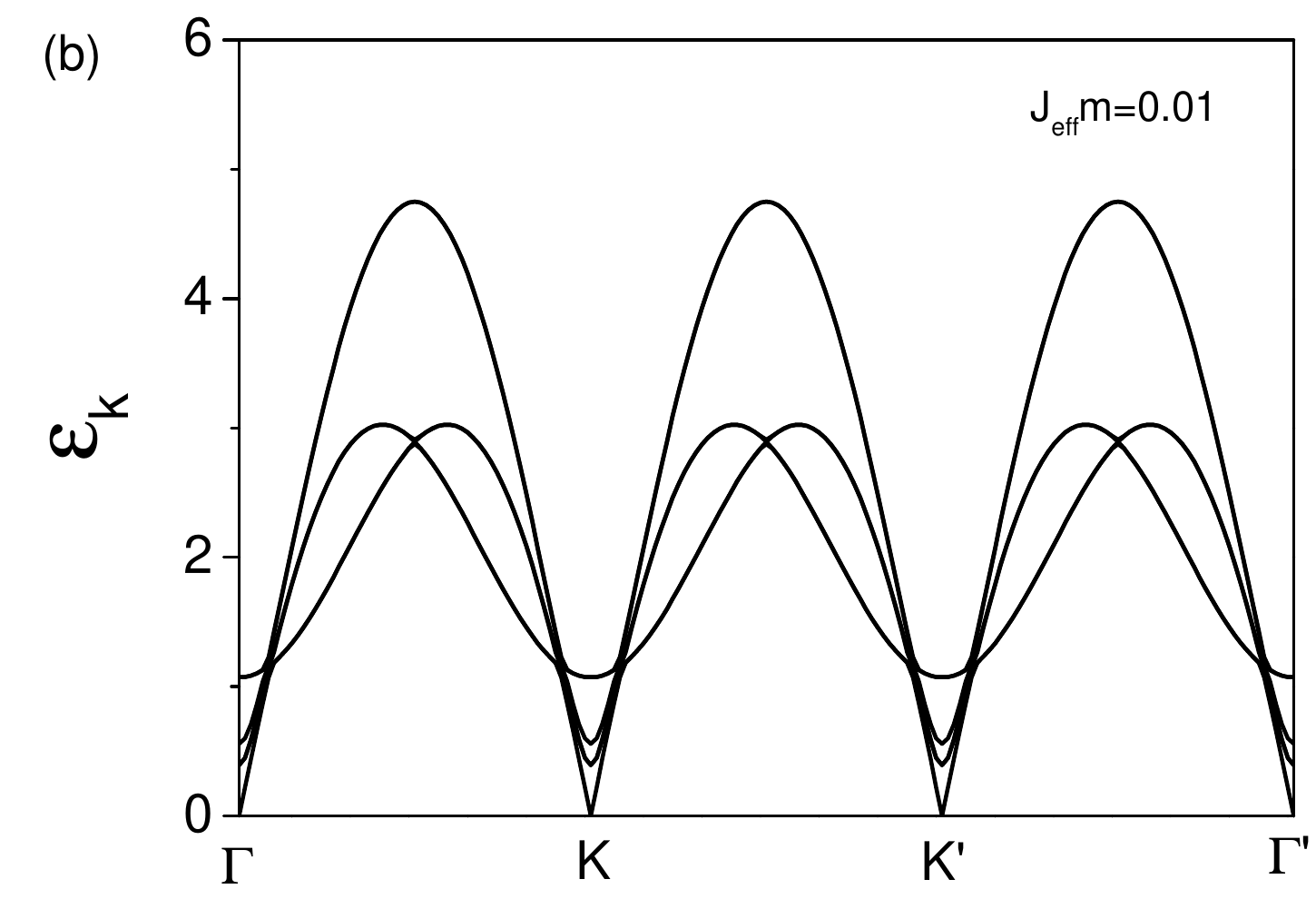}
\includegraphics[width=7cm]{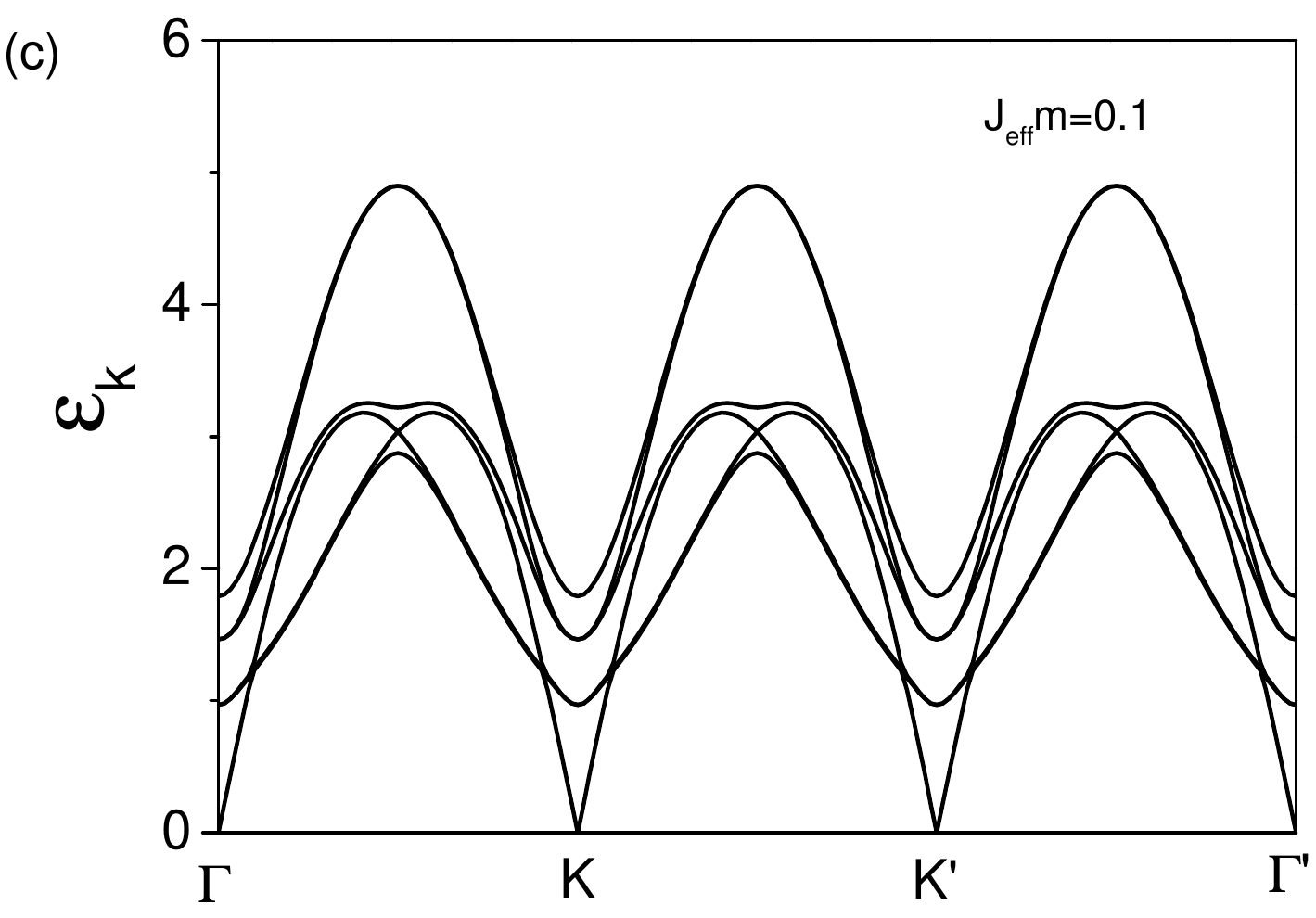}
\includegraphics[width=7cm]{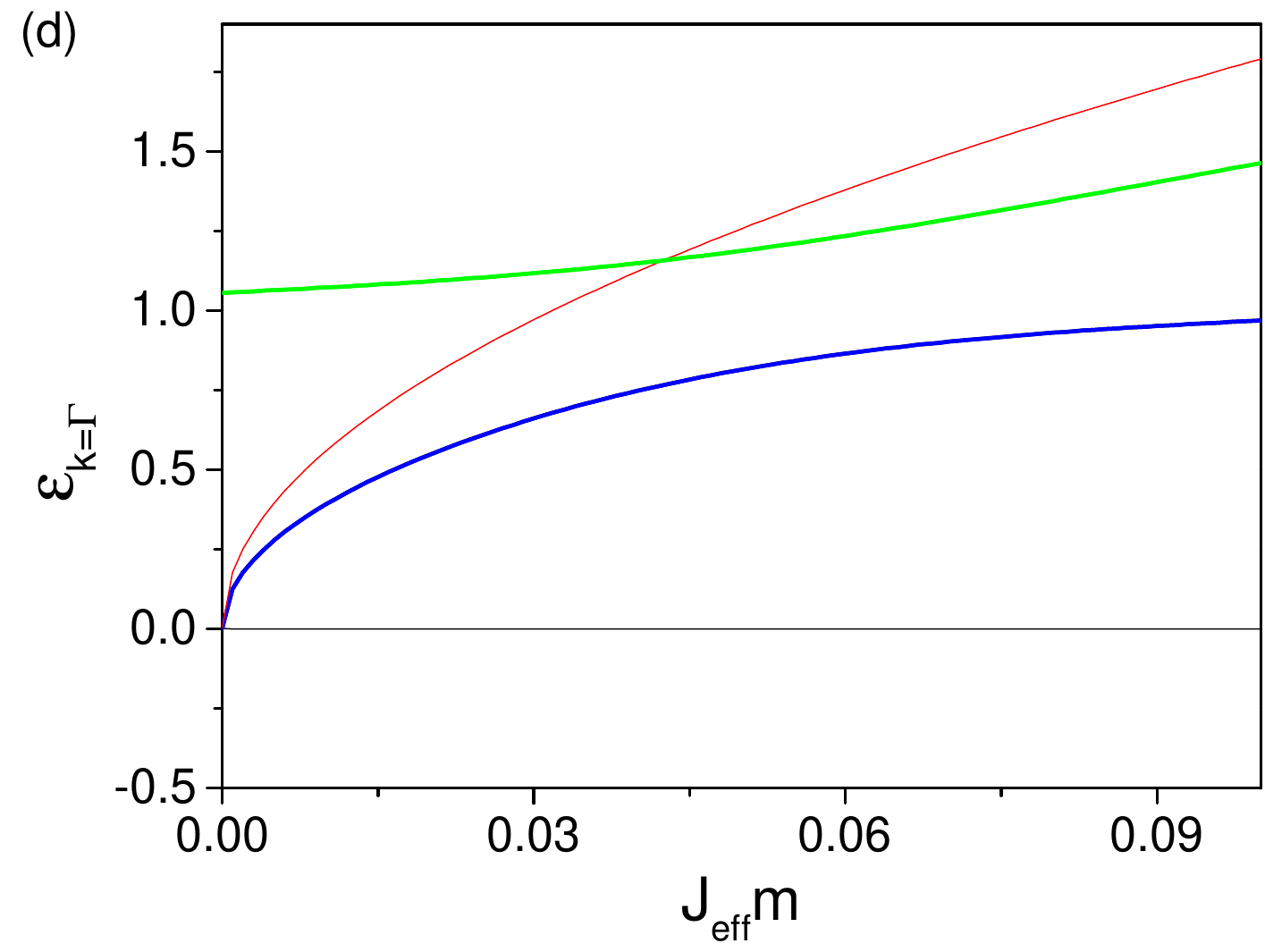}
\caption{The spinon dispersion predicted by the generalized SBMFT along the high symmetry path $\bm{\Gamma}-\mathbf{K}-\mathbf{K'}-\bm{\Gamma}'$ in the Brillouin zone for (a)$\mathrm{J_{eff}m}=0$, (b)$\mathrm{J_{eff}m}=0.01$ and (c)$\mathrm{J_{eff}m}=0.1$. (d)The dependence of the spinon energy at the $\bm{\Gamma}$ point on $\mathrm{J_{eff}m}$. The levels plotted in thick lines are doubly degenerate and those plotted in thin lines are non-degenerate. Here we have set the RVB parameter $A=1$ and choose $B=-0.46$. }
\end{figure}

In the Fermionic RVB theory, the coupling to the magnetic order parameter will gap out the spinon excitation in the low energy limit. However, we find that the spinon spectrum generated from the solution of the mean field equation Eq.(44) and Eq.(46) is always gapless. In Fig.7, we present the spinon dispersion calculated at three different values of $\mathrm{J_{eff}m}$. Among them, the spinon dispersion for $\mathrm{J_{eff}m}=0$ shown in Fig.7a can be obtained by simply folding the dispersion shown in Fig.6 by $2\mathbf{Q}$ and $-2\mathbf{Q}$. Each level is then doubly degenerate. With the increase of $\mathrm{J_{eff}m}$, three out of four gapless modes are gapped out around the $\bm{\Gamma}$ point. The double degeneracy is also lifted at general momentum. In Fig.7d, we present the $\mathrm{J_{eff}m}$ dependence of the gap at the $\bm{\Gamma}$ point for the six modes. We find that the spinon gap increase very quickly with $\mathrm{J_{eff}m}$ for small $\mathrm{J_{eff}m}$(in a square root fashion).

A natural question is then what is the origin of these modes. In particular, what is the origin of the gapless mode. To answer this question, we have computed the contribution of these modes to the  spin fluctuation spectrum of the system. The dynamical spin susceptibility of the system takes the form of a $9\times9$ matrix\cite{Tao20201}
 \begin{equation}
\bm{\chi}_{\mu,\nu}^{i,j}(\mathbf{q},\tau)=-\langle \ T_{\tau} \mathrm{S}^{i}_{\mu}(\mathbf{q},\tau)\  \mathrm{S}^{j}_{\nu}(-\mathbf{q},0) \rangle.
 \end{equation}
Here $\mu,\nu=A,B,C$ denotes the indices of the three magnetic sublattices. $i,j=x,y,z$ denotes the spin components. 
\begin{equation}
\mathrm{S}^{i}_{\mu}(\mathbf{q})=\frac{1}{2}\sum_{\mathbf{k},\alpha,\beta}b^{\dagger}_{\mathbf{k+q},\mu,\alpha}\sigma^{i}_{\alpha,\beta}b_{\mathbf{k},\mu,\beta},
\end{equation}
is the spin density operator in the $i$-th direction and $\mu$-th sublattice. Note that the momentum $\mathbf{q}$ and $\mathbf{k}$ are now defined in the magnetic Brillouin zone. From the Lehmann representation, we can express the dynamical spin susceptibility in terms of the spectral matrix as follows
\begin{equation}
\bm{\chi}^{i,j}_{\mu,\nu}(\mathbf{q},i\omega_{n})=\frac{1}{2\pi}\int d\omega \ \ \frac{\ \mathbf{R}^{i,j}_{\mu,\nu}(\mathbf{q},\omega)\  \ }{\  i\omega_{n}-\omega\  \ },
\end{equation}
in which the spectral matrix is given by
\begin{eqnarray}
\mathbf{R}^{i,j}_{\mu,\nu}(\mathbf{q},\omega)=\frac{1}{Z}\sum_{n,m}e^{-\beta(E_{n}-E_{m})}\langle n| \mathrm{S}^{i}_{\mu}(\mathbf{q})|m\rangle\times\nonumber\\
\langle m |\mathrm{S}^{j}_{\nu}(-\mathbf{q})|n\rangle\times 2\pi \delta(\hbar \omega -(E_{n}-E_{m})).
\end{eqnarray}
Here $Z=\sum_{n}e^{-\beta E_{n}}$ is the partition function of the system. The spectral matrix $\mathbf{R}$ can be extracted from the retarded spin susceptibility matrix by taking its anti-Hermitian part, or
\begin{equation}
\mathbf{R}^{i,j}_{\mu,\nu}(\mathbf{q},\omega)=\frac{1}{2i}\left[ \bm{\chi}^{i,j}_{\mu,\nu}(\mathbf{q},\omega+i0^{+})-\bm{\chi}^{j,i}_{\nu,\mu}(\mathbf{q},\omega-i0^{+})\right].
\end{equation}
It can be shown that the spectral matrix $\mathbf{R}^{i,j}_{\mu,\nu}(\mathbf{q},\omega)$ is positive definite and Hermitian in its index $(i,\mu)$ and $(j,\nu)$. We can thus interpret its eigenvalues and eigenvectors as the spectral weight and polarization vectors of the spin fluctuation at momentum $\mathbf{q}$ and frequency $\omega$.

We now compute $\mathbf{R}^{i,j}_{\mu,\nu}(\mathbf{q},\omega)$ with the generalized SBMFT. We will focus on the contribution from the coherent peak with one spinon resides in the spinon condensate. It is given by
\begin{eqnarray}
\mathbf{R}^{i,j}_{\mu,\nu}(\mathbf{q},\omega)=\sum_{n=1}^{6}w^{i,j}_{\mu,\nu,n} \ \delta(\ \omega-\epsilon^{n}_{\mathbf{q}}\ ),
\end{eqnarray}
with
\begin{eqnarray}
w^{i,j}_{\mu,\nu,n}=\frac{\pi}{4}\sum_{\alpha,\beta,\gamma,\delta=\uparrow\downarrow}\sigma^{i}_{\alpha,\beta} 
\left[  u^{n*}_{\mathbf{q},\mu,\alpha}\bar{b}_{\mu,\beta}+\bar{b}^{*}_{\mu,\alpha}v^{n*}_{\mathbf{q},\mu,\beta} \right]\nonumber\\
\times\sigma^{j}_{\gamma,\delta}\left[  \bar{b}^{*}_{\nu,\gamma} u^{n}_{\mathbf{q},\nu,\delta}+v^{n}_{\mathbf{q},\nu,\gamma} \bar{b}_{\nu,\delta} \right].\nonumber\\
\end{eqnarray}
Here $\bar{b}_{\mu,\alpha}$ denotes the spinon condensate on the $\mu$-th sublattice.  $u^{n}_{\mathbf{q},\mu,\alpha}$ and $v^{n}_{\mathbf{q},\mu,\alpha}$ denote the first and the next 6 components of the $n$-th column of the matrix $U_{\mathbf{q}}$. To better illustrate the polarization character of each mode, it is convenient to project the spectral matrix onto the local coordinate system and define
\begin{eqnarray}
W^{I,J}_{\mu,\nu,n}=\sum_{i,j=x,y,z}\mathbf{e}^{i}_{\mu,I}\ w^{i,j}_{\mu,\nu,n} \ \mathbf{e}^{j}_{\nu,J}.
\end{eqnarray}
Here $I,J=X,Y,Z$ denote the three basis directions of the local coordinate system. $\mathbf{e}_{\mu,I}$ denotes the $I$-th local basis vector in the $\mu$-th sublattice(see Fig.1). 

We have computed the eigenvalues and the eigenvectors of the weight matrix $W^{I,J}_{\mu,\nu,n}$ for each of the six modes shown in Fig.7d. We find that the non-degenerate gapless mode shown as the black thin line has a pure longitudinal character at the $\bm{\Gamma}$ point. The non-degenerate gapped mode shown as the red thin line has a dominate in-plane transverse polarization character. The doubly degenerate gapped mode shown as the blue thick line has a dominate out-of-plane transverse polarization character. The spurious quadratic mode shown as the green thick line is isotropic in spin space and carries a total spin fluctuation weight(defined as the trace of $W$) that is almost two orders of magnitude smaller than that of the three transverse modes. Thus the coupling to the magnetic order will quickly gap out the transverse spinon mode but will leave the longitudinal spinon mode gapless. The gaplessness of the longitudinal mode in the magnetic ordered phase is of course an artifact of the mean field treatment and should be attributed to the lack of spinon rigidity in the SBMFT.

To understand the origin of the spurious quadratic mode, we have computed its contribution to the Boson density fluctuation, whose spectral weight is given by
\begin{eqnarray}
N_{\mu,\nu,n}=\pi\sum_{\alpha,\beta=\uparrow\downarrow}
\left[  u^{n*}_{\mathbf{q},\mu,\alpha}\bar{b}_{\mu,\alpha}+\bar{b}^{*}_{\mu,\alpha}v^{n*}_{\mathbf{q},\mu,\alpha} \right]\nonumber\\
\times\left[  \bar{b}^{*}_{\nu,\beta} u^{n}_{\mathbf{q},\nu,\beta}+v^{n}_{\mathbf{q},\nu,\beta} \bar{b}_{\nu,\beta} \right].
\end{eqnarray} 
We find that the spectral weight of this mode in the density fluctuation channel is comparable to its spectral weight in the spin fluctuation channel. On the other hand, the spectral weight of the three transverse modes in the density fluctuation channel are about 4 orders of magnitude smaller than their spectral weight in the spin fluctuation channel. In particular, the spinon mode with dominate in-plane polarization character has no spectral weight at all in the density fluctuation channel. Since the spurious quadratic mode appears only around the $\mathbf{K}$ and $\mathbf{K}'$, but not the $\bm{\Gamma}$ point of the lattice Brillouin zone when $\mathrm{J_{eff}m}=0$, the density fluctuation in this mode has a nonzero momentum of $\pm2\mathbf{Q}$, which is not allowed by the no double occupancy constraint on the spinon system.

From these analyses, we see that none of the spinon mode predicted by the SBMFT should be interpreted as the Goldstone mode in the magnetic ordered phase, which must be found in the collective motion of the spinon system as a whole. The description of the spin dynamics in the Bosonic RVB theory is thus qualitatively similar to that in the Fermionic RVB theory.

\section{Discussions}
In this work, we find that the SBMFT supplemented with Gutzwiller projection provides an exceedingly accurate description of the ground state of the spin-$\frac{1}{2}$ TLHAF. Furthermore, we find that the Bosonic RVB state derived from such a construction can be deformed continuously into the short ranged RVB state proposed by Anderson and the Dirac spin liquid state proposed by Yunoki and Sorella. However, the SBMFT in its conventional form fails even qualitatively to describe the dynamical behavior of the system in the magnetic ordered phase. More specifically, if we take naively the coherent peak in the two-spinon continuum formed in a background of spinon condensate as the magnon mode, then the SMBFT fails to explain the anisotropy in the magnon velocity and predicts an incorrect number of magnon branches. In addition, the SBMFT predicts a spurious mode with quadratic dispersion around the $\mathbf{K}$ and $\mathbf{K}'$ point and a gapless longitudinal spin fluctuation mode in the magnetic ordered phase. On a more conceptual level, the SBMFT predicts incorrectly the existence of deconfined spinon in the low energy limit in the magnetic ordered phase.

We show that the failure of the SBMFT can be attributed to the following facts. (1)The spinon condensation fails to provide a consistent description of the order parameter manifold of the 120 degree ordered phase. (2)There lacks in the SBMFT the coupling between the uncondensed spinon and the spinon condensate, which breaks both the spin rotational and the translational symmetry. (3)There lacks in the SBMFT the rigidity  that is related to the no double occupancy constraint on the spinon system.  As a result of the first fact, we should not expect the SBMFT to describe the collective spin fluctuation of the magnetic ordered phase. In particular, we should not mistake the coherent peak in the two-spinon continuum predicted by the SBMFT for the magnon excitation. The second fact is responsible for the false prediction of an isotropic 'magnon' velocity and the incorrect number of 'magnon' branches by the SBMFT. The third fact is responsible for the spurious mode around the $\mathbf{K}$ and $\mathbf{K}'$ point and the gaplessness of the longitudinal mode in the magnetic ordered phase.

The first two issues can be resolved by introducing directly the magnetic order parameter into the mean field description. In particular, we find that the coupling with such a symmetry breaking background will quickly gap out the spinon mode corresponding to transverse spin fluctuation. However, without the spinon rigidity related to the no double occupancy constraint, the longitudinal mode is always gapless and the spurious quadratic mode remains around the $\mathbf{K}$ and $\mathbf{K}'$ point. It is rather illuminating here to make a comparison with the corresponding Fermionic RVB theory, in which the magnetic order will gap out spinon excitation totally. The Goldstone mode only appears after we include the RPA correction between the gapped spinons and emerge as the collective motion of the spinon system as a whole. 

We note that the problems with the SBMFT is not restricted to the magnetic ordered phase. For example, the SBMFT predicts that in the paramagnetic phase there should be a quadratic dispersing feature with large spectral weight within the two-spinon continuum around the the $\mathbf{K}$ and $\mathbf{K}'$ point. This feature is the remanent of the spurious mode in the magnetic ordered phase and is an artifact of the mean field theory. It will disappear when we enforce the no double occupancy constraint on the Bosonic spinon. We also note that the problems with the SBMFT is not restricted to the spin-$\frac{1}{2}$ TLHAF. It can be easily shown that the spinon condensation can not describe consistently the order parameter manifold of a general non-collinear magnetic order. In the case of collinear quantum antiferromagnet such as the spin-$\frac{1}{2}$ SLHAF, while it is true that its order parameter manifold can be represented consistently by the spinon condensation, the lack of coupling to the spinon condensation and the lack of spinon rigidity in the SBMFT is still problematic. Especially, the SBMFT predicts incorrectly the existence of gapless spinon excitation in the low energy limit and the gaplessness  of the longitudinal spin fluctuation in the magnetic ordered phase.

In all, we show that the prediction of the SBMFT on the spin dynamics of the quantum antiferromagnet is in general incorrect as a result of its inconsistent description of the symmetry breaking order and its lack of spinon rigidity. The only spectral feature in the low energy limit one should expect in the magnetic ordered phase is the Goldstone mode accompanying the spontaneous symmetry breaking. While our generalized SBMFT can resolve the first problem, it is impossible to resolve the problem related to the spinon rigidity. In fact, the spin dynamics is not the only place where the lack of the spinon rigidity can lead to qualitative failure of the SBMFT. For example, the SBMFT fails to describe the topological property of a chiral spin liquid state. In the SBMFT framework, all spin liquid state can be deformed continuously to the trivial Boson vacuum state by increasing the Bosonic chemical potential, without closing any spectral gap. Thus the topological non-trivialness of the Bosonic RVB state is protected by the no double occupancy constraint. 

While such a Bosonic rigidity is beyond the reach of any mean field treatment, it can be treated faithfully by the Gutzwiller projection technique. In recent years, such a technique has been generalized to the study of the dynamical properties of strongly correlated electron models and received much success\cite{Li,Tao2011,Piazza,Ferrari,Ferrari2,Becca2019,Ido,Imada2020,Tao20202}. In such a variational treatment, one adopt the Gutzwiller projected mean field eigenstates(all those that can be reached by operating the spin density operator $\mathrm{S}^{i}_{\mu}(\mathbf{q})$ on the mean field ground state $|\mathrm{MFG}\rangle$) to form a variational subspace, within which the dynamical behavior of the system is computed. These Gutzwiller projected mean field eigenstates form a non-orthogonal basis of the variational subspace and has a dimension of the order of the system size. To compute the spin fluctuation spectrum one has to solve a generalized eigenvalue problem in such a vartional subspace. It can be shown that the variational spectrum so constructed satisfies the momentum-resolved spin sum rule on the Gutzwiller projected ground state and respects the Goldstone theorem. However, performing variational calculation on Gutzwiller projected Bosonic RVB state is a much more expansive task than that on its Fermionic cousin. A calculation along this line, but under the multi-mode approximation by retaining only the nine basis vectors $\mathrm{S}^{i}_{\mu}(\mathbf{q})|\mathrm{MFG}\rangle$, is now underway.

In conclusion, we show that the Bosonic RVB theory based on Gutzwiller projection of the SBMFT ground state can provide a very accurate description of the ground state of the spin-$\frac{1}{2}$ TLHAF. It is very interesting to see if such a success can be achieved for more general frustrated quantum antiferromagnet. The success of the SBMFT in the description of the ground state of the system implies that it has already build in the essential ingredient for an accurate description of the dynamical behavior of the system. However, we show that it is misleading to interpret a spinon moving in the background of spinon condensation as a magnon mode. In fact, one should in general introduce the magnetic order parameter directly to describe the low energy physics of the magnetic ordered phase. The coupling to such a symmetry breaking background will gap out the spinon in the low energy limit, leaving the magnon, namely the collective fluctuation of the spinon system, as the only spectral feature in the low energy regime. In addition, we show that it is crucial to endow the Bosonic spinon the rigidity to generate the mass of the longitudinal spin fluctuation and to remove the spurious quadratic dispersing mode in the magnetic ordered phase. We show that these understandings are also relevant for more general quantum antiferromagnet.

\begin{acknowledgments}
We acknowledge the support from the grant NSFC 11674391, the Research Funds of Renmin University of China, and the grant National Basic Research Project
2016YFA0300504. We also acknowledge the contribution of Chun Zhang at the early stage of this work.
\end{acknowledgments}

\end{document}